\documentclass[12pt]{amsart}

\usepackage{times}

\usepackage{amsfonts}
\usepackage{amsmath}
\usepackage{amssymb}
\usepackage{amsthm}
\usepackage{graphicx}
\usepackage{ifthen}
\usepackage{color}

\usepackage[colorlinks=true,urlcolor=webblue,linkcolor=webgreen,filecolor=webblue,
citecolor=webgreen,pdfpagemode=UseOutlines,pdfstartview=FitH,pdfpagelayout=OneColumn,
bookmarks=true]{hyperref}

\definecolor{webgreen}{rgb}{0,.5,0}
\definecolor{webblue}{rgb}{0,0,.5}

\newtheorem{theorem}{Theorem}
\newtheorem{lemma}[theorem]{Lemma}

\newtheorem{definition}[theorem]{Definition}

\theoremstyle{remark}

\DeclareMathOperator{\tr}{\textbf{tr}}

\newcommand{\sievefinal}{1}

\newcommand{\prelim}[1]{\ifthenelse{\equal{\sievefinal}{0}}{#1}{}}
\newcommand{\final}[1]{\ifthenelse{\equal{\sievefinal}{1}}{#1}{}}

\DeclareMathAlphabet{\varmathbb}{U}{bbold}{m}{n}
\newcommand{\one}{{\varmathbb 1}}
\renewcommand{\vec}[1]{\mathbf{#1}}

\newcommand{\remove}[1]{}

\newcommand{\C}{\mathbb{C}}

\newcommand{\Z}{\mathbb{Z}}

\newcommand{\U}{\textsf{U}}

\newcommand{\reg}{\textbf{reg}}

\newcommand{\ket}[1]{\left| #1 \right\rangle}
\newcommand{\bra}[1]{\left\langle #1 \right|}
\newcommand{\inner}[2]{\left\langle #1, #2 \right\rangle}
\newcommand{\innerg}[2]{\left\langle #1, #2 \right\rangle_G}

\newcommand{\CG}{\C[G]}
\newcommand{\wg}{\widehat{G}}
\newcommand{\norm}[1]{\left\| #1 \right\|}
\newcommand{\abs}[1]{\left| #1 \right|}

\newcommand{\vsigma}{{\boldsymbol{\sigma}}}

\newcommand{\Root}{{\rm root}}
\newcommand{\calP}{{\mathcal P}}
\newcommand{\planch}{\calP_{\rm planch}}
\newcommand{\Prho}{\calP_\rho}

\newcommand{\prtriv}{P_T^{\{1\}}}
\newcommand{\prh}{P_T^H}
\newcommand{\rhotriv}{\rho_{\{1\}}}
\newcommand{\Plmt}{{\calP_{\lambda \otimes \mu}(\tau)}}
\newcommand{\Pcoll}{\calP^{\rm coll}}

\newcommand{\Ind}[2]{{{\rm Ind}_{#1}^{#2}}}

\begin{document}

\title[On the Impossibility of a Quantum Sieve Algorithm]%
{On the Impossibility of a Quantum Sieve Algorithm for Graph
Isomorphism}

\author{Cristopher Moore}
\email{moore@cs.unm.edu}
\address{University of New Mexico,
Department of Computer Science,
Mail stop: MSC01 1130,
1 University of New Mexico,
Albuquerque, NM 87131-0001, USA}

\address{Santa Fe Institute, 
1399 Hyde Park Road,
Santa Fe, New Mexico 87501,
USA}

\author{Alexander Russell}
\email{acr@cse.uconn.edu}
\address{Department of Computer Science \& Engineering
University of Connecticut
371 Fairfield Rd., U-2155
Storrs, CT 06269, USA}

\author{Piotr \'{S}niady}
\email{piotr.sniady@math.uni.wroc.pl}
\address{Institute of Mathematics, University of Wroclaw, pl.~Grunwaldzki~2/4, 50-384 Wroclaw, Poland}

\date{}

\maketitle

\begin{abstract}
It is known that any quantum algorithm for Graph Isomorphism that
works within the framework of the hidden subgroup problem (HSP) must perform
highly entangled measurements across $\Omega(n \log n)$ coset states.  One of
the only known models for how such a measurement could be carried out
efficiently is Kuperberg's algorithm for the HSP in the dihedral group, in which
quantum states are adaptively combined and measured according to the
decomposition of tensor products into irreducible representations.  This
``quantum sieve'' starts with coset states, and works its way down towards
representations whose probabilities differ depending on, for example, whether
the hidden subgroup is trivial or nontrivial.

In this paper we show that no such approach can produce a po\-ly\-no\-mial-time 
quantum algorithm for Graph Isomorphism.  Specifically, we consider the natural reduction of 
Graph Isomorphism to the HSP over the the wreath product $S_n \wr \Z_2$.  
Using a recently proved bound on the irreducible characters of $S_n$, 
we show that no algorithm in this family can solve Graph Isomorphism in less than
 $e^{\Omega(\sqrt{n})}$ time, no matter what adaptive rule it uses to select and 
combine quantum states.  In particular, algorithms of this type can offer 
essentially no improvement over the best known classical algorithms, which run in time
$e^{O(\sqrt{n \log n})}$. 
\end{abstract}


\section{Introduction}
\label{sec:intro}

The discovery of Shor's and Simon's algorithms began a frenzied charge to uncover the full
algorithmic potential of a general purpose quantum computer. Creative
invocations of the order-finding primitive yielded efficient quantum algorithms
for a number of other number-theoretic
problems~\cite{Hallgren:2002:PTQ,Hallgren:2005:FQA}. As the field matured, these
algorithms were roughly unified under the general framework of the \emph{hidden
subgroup problem}, where one must determine a subgroup $H$ of a group $G$ by
querying an oracle $f: G \rightarrow S$ known to have the property that $f(g) =
f(gh) \Leftrightarrow h \in H$. Solutions to this general problem are the
foundation for almost all known superpolynomial speedups offered by quantum
algorithms over their classical counterparts (see~\cite{Aharonov:2006:PQA} for
an important exception). 

The algorithms of Simon and Shor essentially solve the hidden subgroup problem
on abelian groups, namely $\Z_2^n$ and $\Z_n^*$ respectively.  Since then,
\emph{non-abelian} hidden subgroup problems have received a great deal of
attention (e.g.
\cite{Hallgren:2000:NSR,Grigni:2001:QMA,Friedl:2003:HTO,SODA::MooreRS2004,
Bacon:2005:OMEA,Hallgren:2006:LQC}).  
A major motivation for this work is the fact that we can reduce Graph Isomorphism for
rigid graphs of size $n$ to the case of the hidden subgroup problem over the
symmetric group $S_{2n}$, or more specifically the wreath product $S_n \wr
\Z_2$, where the hidden subgroup is promised to be either trivial or of order two. 
The standard approach to these problems is to prepare ``coset states'' of the
form
\[ \rho_H 
= \frac{1}{|G|} \sum_c \ket{cH}\bra{cH} \enspace,
\]
where $\ket{S}$, for a subset $S \subset G$, denotes the uniform superposition
$(1/\sqrt{|S|}) \sum_{g\in S} \ket{g}$. In the abelian case, one proceeds by computing
the quantum Fourier transform of such coset states, measuring the resulting
states, and appropriately interpreting the results. In the case of the symmetric
group, however, determining $H$ from a quantum measurement of coset states is
far more difficult.  In particular, no \emph{product measurement} (that is, a
measurement which treats each coset state independently) can efficiently
determine a hidden subgroup over $S_n$~\cite{Moore:2005:SGF}; in fact, any
successful measurement must be \emph{entangled} over $\Omega(n \log n)$ coset
states at once~\cite{Hallgren:2006:LQC}.

One of the few proposed methods for building such an entangled measurement 
comes from Kuperberg's algorithm for the hidden subgroup problem in the dihedral
group~\cite{Kuperberg}.  It starts by generating a large number of coset states and
subjecting each one to \emph{weak Fourier sampling}, so that it lies inside a
known irreducible representation.  It then proceeds with an adaptive ``sieve''
process, at each step of which it judiciously selects pairs of states and
measures them in a basis consistent with the \emph{Clebsch-Gordan} decomposition
of their tensor product into irreducible representations.  This sieve continues
until we obtain a state lying in an ``informative'' representation: namely, one
from which information about the hidden subgroup can be easily extracted.  We
can visualize a run of the sieve as a forest, where leaves consist of the initial coset states, each
internal node measures the tensor product of its parents, and the informative
representations lie at the roots.

This approach is especially attractive in cases like Graph Isomorphism,
where all we need to know is whether the hidden subgroup is trivial or
nontrivial.  Specifically, suppose that the hidden subgroup $H$ is promised to
be either the trivial subgroup $\{1\}$ or a conjugate of a known subgroup
$H_0$. Assume further that there is an irreducible representation $\sigma$ of
$G$ with the property that $\sum_{h \in H_0} \sigma(h) = 0$; that is, a
``missing harmonic'' in the sense of~\cite{MooreR:banff}.  In this case, if $H$
is nontrivial then the probability of observing $\sigma$ under weak Fourier
sampling of the coset state $\rho_H$ is zero.  More generally, as we discuss below, 
the irrep $\sigma$ cannot appear at any time in the sieve.  If, on the
other hand, one can guarantee that the sieve \emph{does} observe
$\sigma$ with significant probability when the hidden subgroup is trivial and
the corresponding states are completely mixed, it gives us an algorithm to
distinguish the two cases.  

For example, if we consider the case of the hidden subgroup problem in the
dihedral group $D_n$ where $H$ is either trivial or a conjugate of $H_0=\{1,m\}$
where $m$ is an involution, then the sign representation $\pi$ is a missing
harmonic.  Applying Kuperberg's sieve, we observe $\pi$ with significant
probability after $e^{O(\sqrt{n})}$ steps if $H$ is trivial, while we can never
observe it if $H$ is of order $2$.  A similar approach was applied
to groups of the form $G^n$ by Alagi\'{c} et al.~\cite{AlagicMR06}.

We show here, however, that the hidden subgroup
problem related to Graph Isomorphism cannot be solved efficiently by any
algorithm in this family. Specifically, no matter what adaptive selection rule it uses  
to choose pairs of states to combine and measure, such a sieve cannot distinguish the 
isomorphic and nonisomorphic cases unless it takes $e^{\Omega(\sqrt{n})}$ time 
(and uses this many coset states). 
In comparison, the best known classical algorithms for Graph Isomorphism
run in time $e^{O(\sqrt{n \log n})}$ for general graphs~\cite{Babai:1980:CCL,Babai:1983:CLG} 
and $e^{O(n^{1/3} \log^2 n)}$ for strongly regular graphs~\cite{Spielman}.  Therefore, 
quantum algorithms of this kind can offer no meaningful improvement over their
classical counterparts.

Our proof relies on several ingredients.  First, we give a formal definition of quantum sieve 
algorithms, and we derive a combinatorial description of the probability distributions 
of their observations in the trivial and nontrivial cases.  We then focus on the case where the 
ambient group is a wreath product $G \wr \Z_2$, and show that no information is gained 
until the sieve observes a so-called inhomogeneous representation.  Then, in the case where 
$G=S_n$, we rely on a bound on the characters of the symmetric group proved very recently 
by Rattan and \'Sniady~\cite{RattanSniady06} to show that the total variation distance 
between the trivial and nontrivial cases is at most $e^{-b\sqrt{n}}$ unless the sieve takes 
$e^{a\sqrt{n}}$ time, for some constants $a, b > 0$.

We note that two of the present authors gave this result in conditional form 
in~\cite{MooreR:sieve}, in which they presented a conjectured bound on the characters of $S_n$.  
Indeed, it was this conjecture which inspired the work of~\cite{RattanSniady06}
who proved its weaker version, which, along 
with some additional arguments, allows us to prove the results of~\cite{MooreR:sieve} unconditionally.

We refer the reader to~\cite{Serre77,JamesK1981} 
for an introduction to to  the representation theory of finite groups, and 
in particular of the symmetric group $S_n$.  One fact which we use repeatedly is that the 
\emph{$\tau$-isotypic subspace}, i.e., the subspace of a representation $\sigma$ which 
consists of copies of an irrep $\tau$,  is the image of the projection operator
\[ \Pi_\tau = \frac{1}{|G|} \sum_{g \in G} d_\tau \chi_\tau(g)^*
\sigma(g) \enspace . \]
These projection operators can be combined to create a measurement whose outcomes are 
names of irreducible representations.  
Applying such a measurement to coset states is known as 
weak Fourier sampling; 
we use the term \emph{isotypic sampling} to refer to the more general case of applying 
an arbitrary group action to a multiregister state.

\section{Fourier analysis on finite groups}
\label{sec:representation-theory}

In this section we review the representation theory of finite groups. Our
treatment is primarily for the purposes of setting down notation; we refer the
reader to~\cite{Serre77} for a complete account.
Let $G$ be a finite group. A \emph{representation} $\sigma$ of $G$ is a
homomorphism $\sigma: G \to \U(V)$, where $V$ is a finite-dimensional Hilbert
space and $\U(V)$ is the group of unitary operators on $V$.  The
\emph{dimension} of $\sigma$, denoted $d_\sigma$, is the dimension of the vector
space $V$. 
Fixing a representation $\sigma: G \to \U(V)$, we say that a subspace $W \subset
V$ is \emph{invariant} if $\sigma(g) \cdot W = W$ for all $g \in G$. 
When $\sigma$ has no invariant subspaces other than the trivial subspace $\{
\vec{0} \}$ and $V$ itself, $\sigma$ is said to be \emph{irreducible}. 

If two representations $\sigma$ and $\sigma'$ are the same up to a unitary
change of basis, we say that they are \emph{equivalent}. It is a fact that any
finite group $G$ has a finite number of distinct irreducible representations up
to equivalence and, for a group $G$, we let $\wg$ denote a set of
representations containing exactly one from each equivalence class.  We often
say that each $\sigma \in \wg$ is the \emph{name} of an irreducible
representation, or an \emph{irrep} for short.

The irreps of $G$ give rise to the Fourier transform. Specifically, for a
function $f: G \to \C$ and an element $\sigma \in \wg$, define the \emph{Fourier
transform of $f$ at $\sigma$} to be \[
\hat{f}(\sigma) = \sqrt{\frac{d_\sigma}{|G|}} \sum_{g \in G}
f(g)\sigma(g)\enspace.
\]
The leading coefficients are chosen to the make the transform unitary, so that
it preserves inner products: \[
\langle f_1, f_2\rangle = \sum_g f_1^*(g)f_2(g)
= \sum_{\sigma \in \wg}\tr \!\left(\hat{f_1}(\sigma)^\dagger \cdot \hat{f_2}(\sigma)\right)\enspace.
\]

\remove{ then, there is a nontrivial invariant subspace $W \subset V$ and, as
the inner product $\langle \cdot, \cdot\rangle$ is invariant under the unitary
maps $\sigma(g)$, it is immediate that the dual subspace
$$
W^\perp = \{ \vec{u} \mid \forall \vec{w} \in W, \langle\vec{u},\vec{w}\rangle = 0\}
$$
is also invariant.  Associated with the decomposition $V = W \oplus W^\perp$ is
the natural decomposition of the operators $\sigma(g) = \sigma_W(g) \oplus
\sigma_{W^\perp}(g)$. By repeating this process, any representation $\sigma: G
\to \U(V)$ may } 

If $\sigma$ is \emph{not} irreducible, it can be decomposed into a direct sum of
irreps $\tau_i$, each of which acts on an invariant subspace, and we write
$\sigma \cong \tau_1 \oplus \cdots \oplus \tau_k$. 
In general, a given $\tau$ can appear multiple times in this decomposition, in
the sense that $\sigma$ may have an invariant subspace isomorphic to the direct
sum of $a_\tau$ copies of $\tau$.  In this case $a_\tau$ is called the
\emph{multiplicity} of $\tau$ in the decomposition of $\sigma$.

There is a natural product operation on representations: if $\lambda: G \to
\U(V)$ and $\mu: G \to \U(W)$ are representations of $G$, we may define a new
representation $\lambda \otimes \mu: G \to \U(V \otimes W)$ as 
$(\lambda \otimes \mu)(g): \vec{u} \otimes \vec{v} \mapsto \lambda(g)\vec{u}
\otimes \mu(g)\vec{v}$.  This representation corresponds to the \emph{diagonal
action} of $G$ on $V \otimes W$, in which we apply the same group element to
both parts of the tensor product.  In general, the representation $\lambda
\otimes \mu$ is not irreducible, even when both $\lambda$ and $\mu$ are. This
leads to the \emph{Clebsch-Gordan
  problem}, that of decomposing $\lambda \otimes \mu$ into irreps.

Given a representation $\sigma$ we define the \emph{character} of $\sigma$,
denoted $\chi_\sigma$, to be the trace 
$\chi_\sigma(g) = \tr \sigma(g)$. As the trace of a linear operator is invariant
under conjugation, characters are constant on the conjugacy classes of $G$. 
Characters are a powerful tool for reasoning about the decomposition of
reducible representations. In particular, when
$\sigma = \bigoplus_i \tau_i$ we have $\chi_\sigma = \sum_i \chi_{\tau_i}$ and,
moreover, for
$\sigma, \tau \in \wg$, we have the orthogonality conditions
$$
\innerg{\chi_\sigma}{\chi_\tau} = \frac{1}{|G|} \sum_{g \in G} \chi_\sigma(g)\chi_\tau(g)^*
= \begin{cases} 1 & \sigma = \tau\enspace,\\
  0 & \sigma \neq \tau\enspace. \end{cases}
$$
Therefore, given a representation $\sigma$ and an irrep $\tau$, the multiplicity
$a_\tau$ with which $\tau $ appears in the decomposition of $\sigma$ is
$\innerg{\chi_\tau}{\chi_\sigma}$.  For example, since $\chi_{\lambda \otimes
\mu}(g) = \chi_\lambda(g) \cdot \chi_\mu(g)$, the multiplicity of $\tau$ in the
Clebsch-Gordan decomposition of $\lambda \otimes \mu$ is
$\innerg{\chi_\tau}{\chi_\lambda \chi_\mu}$.

A representation $\sigma$ is said to be \emph{isotypic} if the irreducible
factors appearing in the decomposition are all isomorphic, which is to say that
there is a single nonzero $a_\tau$ in the decomposition above. Any
representation $\sigma$ may be uniquely decomposed into maximal isotypic
subspaces, one for each irrep $\tau$ of $G$; these subspaces are precisely those
spanned by all copies of $\tau$ in $\sigma$. In fact, for each $\tau$ this
subspace is the image of an explicit projection operator $\Pi_\tau$ which can be
written as
\[ \Pi_\tau = \frac{1}{|G|} \sum_{g \in G} d_\tau \chi_\tau(g)^*
\sigma(g) \enspace .  \]
A useful fact is that $\Pi_\tau$ commutes with the group action; that is, for
any $h \in G$ we have 
\begin{multline*} \sigma(h) \Pi_\tau \sigma(h)^\dagger 
 = \frac{1}{|G|} \sum_{g \in G} d_\tau \chi_\tau(g)^* \sigma(hgh^{-1}) 
 = \\ \frac{1}{|G|} \sum_{g \in G} d_\tau \chi_\tau(h^{-1} g h)^* \sigma(g) 
 = 
\frac{1}{|G|} \sum_{g \in G} d_\tau \chi_\tau(g)^* \sigma(g) 
= \Pi_\tau \enspace .
\end{multline*}

Our algorithms will perform measurements which project into these maximal
isotypic subspaces and observe the resulting irrep name $\tau$.  For the
particular case of coset states, this measurement is called \emph{weak Fourier
sampling} in the literature; however, since we are interested in a more general
process which in fact performs a kind of strong multiregister sampling on the
original coset states, we will use the term \emph{isotypic sampling} instead.
Finally, we discuss the structure of a specific representation, the
\emph{(right) regular} representation $\reg$, which plays an important role in
the analysis below. $\reg$ is given by the permutation action of $G$ on itself. 
Specifically, let $\CG$ be the \emph{group algebra} of $G$; this is the
$|G|$-dimensional vector space of formal sums
$$
\Bigl\{ \sum_g \alpha_g \cdot g \mid \alpha_g \in \C \Bigr\} \enspace .
$$
(Note that $\CG$ is precisely the Hilbert space of a single register containing
a superposition of group elements.) Then $\reg$ is the representation $\reg: G
\to \U(\CG)$ given by linearly extending right multiplication, $\reg(g): h
\mapsto hg$.  It is not hard to see that its character $\chi_\reg$ is given by
$$
\chi_\reg(g) = \begin{cases} |G| & g = 1\enspace,\\
  0 & g \neq 1 \enspace ,
\end{cases}
$$
in which case we have $\innerg{\chi_\reg}{\chi_\sigma} = d_\sigma$ for each
$\sigma \in \wg$.  Thus $\reg$ contains $d_\sigma$ copies of each irrep $\sigma
\in \wg$, and counting dimensions on each side of this decomposition implies
\begin{equation}
\label{eq:planch}
|G| = \sum_{\sigma \in \wg} d_\sigma^2 \enspace.
\end{equation}
This equation suggests a natural probability distribution on $\wg$, the
\emph{Plan\-che\-rel distribution}, which assigns to each irrep $\sigma$ the
probability $\planch^G(\sigma) = d_\sigma^2 / |G|$.  This is simply the
dimensionwise fraction of $\CG$ consisting of copies of $\sigma$; indeed, if we
perform isotypic sampling on the completely mixed state on $\CG$, or
equivalently the coset state where the hidden subgroup is trivial, we observe
exactly this distribution.

In general, we can consider subspaces of $\CG$ that are invariant under
left multiplication, right multiplication, or both; these subspaces are called
\emph{left-}, \emph{right-}, or \emph{bi-invariant} respectively.  For each
$\sigma \in \wg$, the maximal $\sigma$-isotypic subspace is a
$d_\sigma^2$-dimensional bi-invariant subspace; it can be broken up further into
$d_\sigma$ $d_\sigma$-dimensional left-invariant subspaces, or (transversely)
$d_\sigma$ $d_\sigma$-dimensional right-invariant subspaces.  However, this
decomposition is not unique. If $\sigma$ acts on a vector space $V$, then
choosing an orthonormal basis for $V$ allows us to view $\sigma(g)$ as a
$d_\sigma \times d_\sigma$ matrix. Then $\sigma$ acts on the
$d_\sigma^2$-dimensional space of such matrices by left or right multiplication,
and the columns and rows correspond to left- and right-invariant spaces
respectively.

\section{Clebsch-Gordan sieves}
\label{sec:sieves}

Consider the hidden subgroup problem over a group $G$ with the added promise
that the hidden subgroup $H$ is either the trivial subgroup, or a conjugate of
some fixed nontrivial subgroup $H_0$.  We shall consider sieve algorithms for
this problem that proceed as follows: 

\smallskip 
1. The oracle is used to generate $\ell=\ell(n)$ coset states $\rho_H$, each
of which is subjected to weak Fourier sampling. 
This results in a set of states $\rho_i$, where $\rho_i$ is a mixed state known
to lie in the $\sigma_i$-isotypic subspace of $\C[G]$ for some irrep
$\sigma_i$.  

\smallskip 
2. The following \emph{combine-and-measure} procedure is then
repeated as many times as we like. 
Two states $\rho_i$ and $\rho_j$ in the set are selected according to an
arbitrary adaptive rule that may depend on the entire history of the computation
(in existing algorithms of this type, this selection in fact depends only on the
irreps $\sigma_i$ and $\sigma_j$ in which they lie).  We then perform isotypic
sampling on their tensor product $\rho_i \otimes \rho_j$: that is, we apply a
measurement operator which observes an irrep $\sigma$ in the Clebsch-Gordan
decomposition of $\sigma_i \otimes \sigma_j$ (see~\cite{Kuperberg}
or~\cite{MooreR:banff} for how this measurement can actually be carried out by
applying the diagonal action).  This measurement destroys $\rho_i$ and $\rho_j$,
and results in a new mixed state $\rho$ which lies in the maximal
$\sigma$-isotypic subspace; we add this new state to the set. 

\smallskip 
3. Finally, depending on the sequence of observations obtained throughout this process, 
the algorithm guesses the hidden subgroup.

\smallskip
We set down some notation to discuss the result of applying such an algorithm.
Fixing a group $G$ and a subgroup $H$, let $A$ be a sieve algorithm which
initially generates $\ell$ coset states.  As a bookkeeping tool, we will
describe intermediate states of $A$'s progress as a \emph{forest of labeled
binary trees}.  Throughout, we will maintain the invariant that the roots of the
trees in this forest correspond to the current set of states available to the
algorithm.  

Initially, the state of the algorithm consists of a forest consisting of $\ell$
single-node trees, each of which is labeled with the irrep name $\sigma_i$ that
resulted from weak Fourier sampling a coset state, and is
associated with the resulting state $\rho_i$.  Then, each combine-and-measure
step selects two root nodes, $r_1$ and $r_2$, and applies isotypic sampling to
the tensor product of their states.  We associate the resulting state $\rho$
with a new root node $r$, and place the nodes $r_1$ and $r_2$ below it as its
children.  We label this new node with the irrep name $\sigma$ observed in this measurement.

Thus, every node of the forest corresponds to a state that existed at some point
during the algorithm, and each node $i$ is labeled with the name of the irrep
$\sigma_i$ observed in the isotypic measurement performed when that node was
created.  We call the resulting labeled forest the \emph{transcript} of the
algorithm: note that this transcript contains all the information the algorithm
may use to determine the hidden subgroup. 

We make several observations about algorithms of this type.  First, it is easy
to see that nothing is gained by combining $t > 2$ states at a time; we can
simulate this with an algorithm which builds a binary tree with $t$ leaves, and
which ignores the results of all its measurements except the one at the root.

Second, the algorithm maintains the following kind of symmetry under the action
of the subgroup $H$.  Suppose we have a representation $\sigma$ acting on a
Hilbert space $V$.  Given a subgroup $H$, we say that a state $\psi \in V$ is
\emph{$H$-invariant} if $\sigma(h) \cdot \psi = \psi$ for all $h \in H$. 
Similarly, given a mixed state $\rho$, we say that $\rho$ is $H$-invariant if
$\sigma(h) \cdot \rho \cdot \sigma(h)^\dagger = \rho$ or, equivalently, if
$\sigma(h)$ and $\rho$ commute.  For instance, the coset state $\rho_H$ is
$H$-invariant under the right regular representation, since right-multiplying by
any $h \in H$ preserves each left coset $cH$.  Now, suppose that $\rho_1$ and
$\rho_2$ are $H$-invariant; clearly $\rho_1 \otimes \rho_2$ is $H$-invariant
under the diagonal action, and performing isotypic sampling preserves
$H$-invariance since $\Pi_\tau$ commutes with the action of any group element. 
Thus the states produced by the algorithm are $H$-invariant throughout.

Third, it is important to note that while at each stage we observe only an irrep
name, rather than a basis vector inside that representation, by iterating this
process the sieve algorithm actually performs a kind of \emph{strong
multiregister Fourier sampling} on the original set of coset states.  For
instance, in the dihedral group, suppose that performing weak Fourier sampling
on two coset states results in the two-dimensional irreps $\sigma_j$ and
$\sigma_k$, and that we then observe the irrep $\sigma_{j+k}$ under isotypic
sampling of their tensor product.  We now know that the original coset states
were in fact confined to a particular subspace, spanned by two entangled pairs
of basis vectors.
Finally, we note that the states produced by a sieve algorithm are quite
different from coset states.  In particular, they belong not to a maximal
isotypic subspace of $\CG$, but to a (typically much higher-dimensional) 
\emph{non-maximal} isotypic subspace of $\CG^{\otimes \ell}$, where $\ell$ is
the number of coset states feeding into that state (i.e., the number of leaves
of the corresponding tree).  Moreover, they have more symmetry than coset
states, since each isotypic measurement implies a symmetry with respect to the
diagonal action on the set of leaves descended from the corresponding internal
node.  In the next sections we will show how these states can be written in
terms of projection operators applied to this high-dimensional space.

\remove{
One special case in which sieve algorithms are attractive is when the nontrivial
subgroup $H_0$ possesses a \emph{missing harmonic}, i.e., an irrep $\tau$ such
that $\sum_{h \in H_0} \tau(h) = 0$.  Then as shown in~\cite{MooreR:banff}, we
can never observe $\tau$ at any point in the sieve; and so if we do observe it,
we know that the hidden subgroup is trivial.  This is how, for instance, the
sieve algorithm for groups of the form $G^n$ in~\cite{AlagicMR06} works. 
However, the negative results we present here do not rely on the notion of a
missing harmonic. }

\section{Observed distributions for fixed to\-po\-lo\-gies}
\label{sec:distributions}

In general, the probability distributions arising from the combine-and-measure
steps of a sieve algorithm depend on both the hidden subgroup and the entire
history of previous measurements and observations (that is, the labeled forest,
or transcript, describing the algorithm's history thus far). In this section and
the next, we focus on the probability distribution induced by a \emph{fixed
forest topology} and subgroup $H$.  We can think of this either as the
probability distribution conditioned on the forest topology, or as the
distribution of transcripts produced by some \emph{non-adaptive} sieve
algorithm, which chooses which states it will combine and measure ahead of time.
We will show
that for all forest topologies of
sufficiently small size, the induced distributions on irrep labels fail to
distinguish trivial and nontrivial subgroups.  Then, in Section~\ref{sec:main},
we will complete the argument for adaptive algorithms.
Clearly, in this non-adaptive case the distributions of irrep labels associated
with different trees in the  forest are independent.  Therefore, we can focus on
the distribution of labels for a specific tree. At the leaves, the labels are
independent and identically distributed according to the distribution resulting
from weak Fourier sampling a coset state~\cite{Hallgren:2000:NSR}. 
However, as we move inside the tree and condition on the irrep labels observed
previously, the resulting distributions are quite different from this initial
one.  To calculate the resulting joint probability distribution, we need to
define projection operators acting on $\CG^{\otimes \ell}$ corresponding to the
isotypic measurement at each node.

First, note that the coset state $\rho_H$ can be written in the following
convenient form: \[ \rho_H 
= \frac{1}{|G|} \sum_c \ket{cH}\bra{cH} 
= \frac{1}{|G|} \sum_{h \in H} \reg(h) \]
where $\reg$ is the \emph{right regular representation}: 
that is, $\rho_H$ is proportional to the projection
operator which right-multiplies by a random element of $H$, \[ \Pi_H =
\frac{1}{|H|} \sum_{h \in H} \reg(h) \enspace . \]
If $H$ is trivial, $\rho_H$ is the completely mixed state
$\rhotriv = (1/|G|) \one$.  On the other hand, if $H=\{1,m\}$ for an involution
$m$, then $\rho_H = (2/|G|) \Pi_H$, where $\Pi_H$ is the projection operator
\[ \Pi_H = \frac{1}{2} (1+\reg(m)) \enspace . \]

Now consider the tensor product of $\ell$ ``registers,'' each containing a coset
state.  Given a linear operator $M$ on $\C[G]$ and a subset $I \subseteq [\ell]
= \{1,\ldots,\ell\}$, let $M^I$ 
denote the operator on $\C[G^\ell] \cong \C[G]^{\otimes \ell}$ which applies $M$
to the registers in $I$ and leaves the other registers unchanged.  Then the
mixed state consisting of $\ell$ independent coset states is 
$\rho_H^{\otimes \ell} = (2/|G|)^\ell \Pi_H^{\otimes \ell}$, where
\begin{equation}
\label{eq:h} 
\Pi_H^{\otimes \ell}
= \frac{1}{2^\ell} \prod_{j=1}^\ell (1+\reg(m)^{\{j\}}) 
= \frac{1}{2^\ell} \sum_{I \subseteq [\ell]} \reg(m)^I 
\enspace. 
\end{equation}
Note the sum over subsets of registers, a theme which has appeared repeatedly in
discussions of multiregister Fourier
sampling~\cite{Regev:2002:QCL,BaconCvD,Hallgren:2006:LQC,Kuperberg,MooreR:banff,MooreR:part2}.
Now consider a tree $T$ with $\ell$ leaves corresponding to the $\ell$ initial
registers, and $k$ nodes including the leaves.  We represent this tree as a
set system, in which each node $i$ is associated with the subset $I_i \subseteq
[\ell]$ of leaves descended from it.  In particular, $I_\Root = [\ell]$ and
$I_j=\{ j \}$ for each leaf $j$.  

Performing isotypic sampling at a node $i$ corresponds to applying the diagonal
action to its children (or in terms of the algorithm, its parents) and
inductively to the registers in $I_i$: that is, we multiply each register in
$I_i$ by the same element $g$ and leave the others fixed.   If $\sigma_i$ is the
irrep label observed at that node, let us denote its character and dimension by
$\chi_i$ and $d_i$ respectively, rather than the more cumbersome
$\chi_{\sigma_i}$ and $d_{\sigma_i}$.  Then the projection operator
corresponding to this observation is \begin{equation}
\label{eq:node}
 \Pi^T_i = \frac{1}{|G|} \sum_{g \in G} d_i \chi_i(g)^* \,\reg(g)^{I_i} 
 \enspace . 
\end{equation} 
Now consider a transcript of the sieve process which results in observing a set
of irrep labels $\vsigma = \{\sigma_i\}$ on the internal nodes of the tree.  The
projection operator associated with this outcome is \begin{equation}
\label{eq:pi_t}
\Pi^T[\vsigma] = \prod_{i=1}^k \Pi^T_i \enspace . 
\end{equation}
We will abbreviate this as $\Pi^T$ whenever the context is clear.  Note that the
various $\Pi^T_i$ in the product~\eqref{eq:pi_t} pairwise commute, since for any
two nodes $i, j$ either $I_i$ and $I_{j}$ are disjoint, or one is contained in
the other.  In the former case $a^{I_i}$ and $b^{I_{j}}$ for all $a,b$.  In the
latter case, say if $I_i \subset I_{j}$, we have $a^{I_i} b^{I_{j}} = b^{I_{j}}
(b^{-1} a b)^{I_i}$, and since $\chi_i(b^{-1} a b) = \chi_i(a)$ it follows
from~\eqref{eq:node} that $\Pi^T_i \Pi^T_{j} = \Pi^T_{j} \Pi^T_i$.

Given a tree $T$ with $k$ nodes, we write $\prtriv[\vsigma]$ for the probability
that we observe the set of irrep labels $\vsigma=\{\sigma_i\}$ in the case where
the hidden subgroup is trivial.  Since the tensor product of coset states is
then the completely mixed state in $\C[G^\ell]$, this is simply the
dimensionwise fraction of $\C[G^\ell]$ consisting of the image of $\Pi^T$, or \[
\prtriv[\vsigma] = \frac{1}{|G|^\ell} \tr \Pi^T \enspace . 
\]
Moreover, since measuring a completely mixed state results in the completely
mixed state in the observed subspace, each state produced by the algorithm is
completely mixed in the image of $\Pi^T$.  In particular, if the irrep label at
the root of a tree is $\sigma$, the corresponding state consists of a classical
mixture across some number of copies of $\sigma$, in each of which it is
completely mixed.  Thus, when combining two parent states with irrep labels
$\lambda$ and $\mu$, we observe each irrep $\tau$ with probability equal to the
dimensionwise fraction of $\lambda \otimes \mu$ consisting of copies of $\tau$,
namely \begin{equation}
\label{eq:ptau}
 \Plmt = \frac{d_\tau}{d_\lambda d_\mu} \inner{\chi_\tau}{\chi_\lambda \chi_\mu}_G
\end{equation}
(recall that 
$\inner{\chi_\tau}{\chi_\rho}_G = (1/|G|) \sum_{g \in G} \chi_\tau(g) \chi_\rho^*(g)$ 
is the multiplicity of $\tau$ in the decomposition of a representation $\rho$ into irreducibles).    
We will refer to this as the \emph{natural distribution} in $\lambda \otimes \mu$.

Now let us consider the case where the hidden subgroup is nontrivial.  Since the
mixed state $\rho_{H^\ell}$ can be thought of as a pure state chosen randomly
from the image of $\Pi_H^{\otimes \ell}$, the probability of observing a set of
irrep labels $\vsigma$ in this case is
\[ \prh[\vsigma] = \frac{\tr \Pi^T \Pi_H^{\otimes \ell}}{\tr \Pi_H^{\otimes
\ell}} = \frac{2^\ell}{|G|^\ell} \tr \Pi^T \Pi_H^{\otimes \ell} \] 
where we use the fact that $\tr \Pi_H^{\otimes \ell} = [G:H]^\ell =
(|G|/2)^\ell$.  Below we abbreviate these distributions as $\prtriv$ and $\prh$
whenever the context is clear.  Our goal is to show that, until the tree $T$ is
deep enough, these two distributions are extremely close, so that 
the algorithm fails to distinguish subgroups of the form $\{1,m\}$ from the
trivial subgroup.

Now let us derive explicit expressions for $\prtriv$ and $\prh$.  First,
we fix some additional notation.  Given an assignment of group elements
$\{a_i\}$ to the nodes, for each leaf $j$ we let  $\prod_{i \leadsto j} a_i$
denote the product of the elements along the path from the root to $j$: \[
\prod_{i \leadsto j} a_i = \prod_{i:j \in I_i} a_i \]
where the product is taken in order from to the root to the leaf.  Then
using~\eqref{eq:node} and~\eqref{eq:pi_t} we can write
\begin{equation}
\label{eq:piT}
\Pi^T = \frac{1}{|G|^k} \sum_{\{a_i\}} d_i \chi_i(a_i)^* 
\bigotimes_{j=1}^\ell \reg\!\left( \prod_{i \leadsto j} a_i \right)^{\!\{i\}} \enspace . 
\end{equation}
We say that an assignment $\{a_i\}$ is \emph{trivial} if $\prod_{i \leadsto j}
a_i = 1$ for every leaf $j$.  Then, since $\tr \reg(g) = \chi_\reg(g) = |G|$ if
$g=1$ and $0$ otherwise, we have 
\begin{equation}
\label{eq:prtriv}
\prtriv = \frac{1}{|G|^\ell} \tr \Pi^T
= \frac{1}{|G|^k} \sum_{\{a_i\}\; \text{trivial}} \prod_{i=1}^k d_i \chi_i(a_i)^*
\enspace . 
\end{equation}
To get a sense of how this expression scales, note that the particular trivial
assignment where $a_i = 1$ for all $i$ contributes $\prod_{i=1}^k d_i^2 / |G| =
\prod_i \planch(\sigma_i)$, as if the $\sigma_i$ were independent and
Plancherel-distributed.

Now consider $\prh$.  Combining~\eqref{eq:h} with~\eqref{eq:piT} gives
the following expression for $\Pi^T \Pi_H^{\otimes \ell}$: 
\begin{equation}
\label{eq:prh}
\Pi^T \Pi_H^{\otimes \ell} = \frac{1}{2^\ell |G|^k} \sum_{\{a_i\}} d_i
\chi_i(a_i)^* \bigotimes_{j=1}^\ell \reg\!\left( \left( \prod_{i \leadsto j} a_i
\right) (1 + m) \right)^{\!\{i\}} \enspace . 
\end{equation}
We say that an assignment $\{a_i\}$ is  \emph{legal} if $\prod_{i \leadsto j}
a_i \in \{1,m\}$ for every leaf $j$.  Then the trace of the term corresponding
to $\{a_i\}$ is $|G|^\ell$ if $\{a_i\}$ is legal, and is $0$ otherwise, and
analogous to~\eqref{eq:prtriv} we have 
\begin{equation}
\label{eq:prh2}
\prh = \frac{2^\ell}{|G|^\ell} \tr \Pi^T \Pi_H^{\otimes \ell} 
= \frac{1}{|G|^k} \sum_{\{a_i\} \;\text{legal}} \prod_{i=1}^k d_i \chi_i(a_i)^* \enspace . 
\end{equation}
Thus these two distributions differ exactly by the terms corresponding to
assignments which are legal but nontrivial.  Our main result will depend on the
fact that for most $\vsigma$ these terms are identically zero, in which case
$\prh$ and $\prtriv$ coincide.

\section{The importance of being homogeneous}
\label{sec:wreath}

For any group $G$, the wreath product $G \wr \Z_2$ is the semidirect product $(G
\times G) \rtimes \Z_2$, where we extend $G \times G$ by an involution which
exchanges the two copies of $G$.  Thus the elements $((\alpha,\beta),0)$ form a
normal subgroup $K \cong G \times G$ of index 2, and the elements
$((\alpha,\beta),1)$ form its nontrivial coset.  We will call these elements
``non-flips'' and ``flips,'' respectively.  
The Graph Isomorphism problem reduces to the hidden subgroup problem on $S_n \wr
\Z_2$ in the following natural way.  We consider the disjoint union of the two
graphs, and consider permutations of their $2n$ vertices.  Then $S_n \wr \Z_2$
is the subgroup of $S_{2n}$ which either maps each graph onto itself (the
non-flips) or exchanges the two graphs (the flips).  We assume for simplicity
that the graphs are rigid.  Then if they are nonisomorphic, the hidden subgroup
is trivial; if they are isomorphic, $H=\{1,m\}$ where $m$ is a flip of the form
$((\alpha,\alpha^{-1}),1)$, where $\alpha$ is the permutation describing the
isomorphism between them.  

For any group $G$, the irreps of $G \wr \Z_2$ can be written in a simple way in
terms of the irreps of $G$.  It is useful to construct them by \emph{inducing}
upward from the irreps of $K \cong G \times G$ (see~\cite{Serre77} for the
definition of an induced representation).  First, each irrep of $K$ is the
tensor product $\lambda \otimes \mu$ of two irreps of $G$.  Inducing this irrep
from $K$ up to $G$ gives a representation \[ \sigma_{\{\lambda,\mu\}} =
\Ind{K}{G} (\lambda \otimes \mu) \]
of dimension $2 d_\lambda d_\mu$.  If $\lambda \not\cong \mu$, then this is
irreducible, and $\sigma_{\{\lambda,\mu\}} \cong \sigma_{\{\mu,\lambda\}}$
(hence the notation).  We call these irreps \emph{inhomogeneous}.  Their
characters are given by
\begin{equation}
\label{eq:inhomchar}
\chi_{\{\lambda,\mu\}}((\alpha,\beta),t) 
= \begin{cases} 
  \chi_\lambda(\alpha) \chi_\mu(\beta) + \chi_\mu(\alpha) \chi_\lambda(\beta) &
\text{if}\;t=0 \\ 0 & \text{if}\;t=1 
\end{cases} \enspace .
\end{equation}
In particular, the character of an inhomogeneous irrep is zero at any flip.

On the other hand, if $\lambda \cong \mu$, then $\sigma_{\{\lambda,\lambda\}}$
decomposes into two irreps of dimension $d_\lambda^2$, which we denote
$\sigma_{\{\lambda,\lambda\}}^+$ and $\sigma_{\{\lambda,\lambda\}}^-$.  We call
these irreps \emph{homogeneous}.  Their characters are given by 
\begin{equation}
\label{eq:homchar}
\chi_{\{\lambda,\lambda\}}^\pm((\alpha,\beta),t)
= \begin{cases} 
  \chi_\lambda(\alpha) \chi_\lambda(\beta) & \text{if}\;t = 0 \\
  \pm \chi_\lambda(\alpha\beta) & \text{if}\;t=1 
  \end{cases}  \enspace . 
\end{equation}
In the next section, we will show that sieve algorithms obtain precisely zero
information that distinguishes hidden subgroups of the form $\{1,m\}$ from the
trivial subgroup until it observes at least one homogeneous representation.

Suppose that the irrep labels $\vsigma = \{ \sigma_i \}$ observed during a run
of the sieve algorithm consist entirely of inhomogeneous irreps of $G \wr \Z_2$.
 Since the irreps have zero character at any flip, the only trivial or legal
assignments $\{ a_i \}$ that contribute to the sums~\eqref{eq:prtriv}
and~\eqref{eq:prh2} are those where each $a_i$ is a non-flip, i.e., is contained
in the subgroup $K \cong G \times G$.  But the product of any string of such
elements is also contained in $K$, so if this product is in $H=\{1,m\}$ where $m
\notin K$, it is equal to $1$.  Thus any legal assignment of this kind is
trivial, the sums~\eqref{eq:prtriv} and~\eqref{eq:prh2} coincide, and the
probability of observing $\vsigma$ is the same in the trivial and nontrivial
cases. That is, so long as every $\sigma_i$ in $\vsigma$ is inhomogeneous,
\begin{equation}
\label{eq:equality}
P^H_T[\vsigma] = P^{\{1\}}_T[\vsigma] \enspace .
\end{equation}
\remove{
This implies the following bound on the total variation distance between the two
distributions, 
\begin{equation}
\label{eq:same}
 \frac{1}{2} \norm{\prh-\prtriv}_1 
 \le \Pr[ \text{at least one homogeneous irrep $\sigma$ has been observed in $T$} \mid H=\{1\} ] 
 \enspace . 
\end{equation}
}
Our strategy will be to show that observing even a single homogeneous irrep is
unlikely, unless the tree generated by the sieve algorithm is quite large. 
Moreover, because the two distributions coincide unless this occurs, it suffices
to show that this is unlikely in the case where $H$ is trivial.
Now, it is easy to see that the probability of observing a given representation
in $G \wr \Z_2$, under either the Plancherel distribution or a natural distribution, 
factorizes neatly into the probabilities that we observe the 
corresponding pair of irreps, in either order, in a pair of similar experiments
in $G$.  First, the Plancherel measure of an inhomogeneous irrep
$\sigma_{\{\lambda,\mu\}}$ is 
\begin{equation}
\label{eq:planch-wreath-inhom}
 \planch^{G \wr \Z_2}(\sigma_{\{\lambda,\mu\}})
= \frac{(2 d_\lambda d_\mu)^2}{2 |G|^2}
= 2 \,\planch^G(\lambda) \,\planch^G(\mu) \enspace . 
\end{equation}
Similarly, the probability that we observe a homogeneous irrep
$\sigma^\pm_{\{\lambda,\lambda\}}$ is the probability of observing $\lambda$
twice under the Plancherel distribution in $G$, in which case the sign $\pm$ is
chosen uniformly: \begin{equation}
\label{eq:planch-wreath-hom}
 \planch^{G \wr \Z_2}(\sigma^\pm_{\{\lambda,\lambda\}})
= \frac{d_\lambda^4}{|G|^2} 
= \planch^G(\lambda)^2 \enspace .
\end{equation}

Now consider the natural distribution in the tensor product of two inhomogeneous
irreps $\sigma_{\{\lambda,\lambda'\}}$ and $\sigma_{\{\mu,\mu'\}}$.  The
multiplicity of a given homogeneous irrep $\sigma^\pm_{\{\tau,\tau\}}$ in this
tensor product, equal to 
\[
\inner{\chi_{\{\tau,\tau\}}^\pm}{\chi_{\{\lambda,\lambda'\}}
\chi_{\{\mu,\mu'\}}}_{G \wr \Z_2} \enspace,
\]
factorizes as follows
\[
 \frac{\innerg{\chi_\tau}{\chi_\lambda \chi_\mu} \innerg{\chi_\tau}{\chi_{\lambda'} \chi_{\mu'}}}{2} 
+ \frac{\innerg{\chi_\tau}{\chi_\lambda \chi_{\mu'}} \innerg{\chi_\tau}{\chi_{\lambda'} \chi_\mu}}{2} \enspace .
\]
Thus the probability of observing either $\sigma_{\{\tau,\tau\}}^+$ or
$\sigma_{\{\tau,\tau\}}^-$ under the natural distribution is 
\begin{equation}
\label{eq:prob-hom}
\calP_{\sigma_{\{\lambda,\lambda'\}} \otimes \sigma_{\{\mu,\mu'\}}}(\sigma_{\{\tau,\tau\}}^\pm)
= \frac{1}{2} \left( \calP_{\lambda \otimes \mu}(\tau) \calP_{\lambda' \otimes \mu'}(\tau) 
+ \calP_{\lambda \otimes \mu'}(\tau) \calP_{\lambda' \otimes \mu}(\tau) \right) \enspace . 
\end{equation}
In other words, the probability of observing a homogeneous irrep of $G \wr \Z_2$
is the probability of observing the same irrep in two natural distributions on
$G$.  Let us denote the probability that we observe the same irrep in the
natural distributions in $\lambda \otimes \mu$ and $\lambda' \otimes
\mu'$---that is, that these two distributions collide---as 
\[ \Pcoll_{\lambda \otimes \mu, \lambda' \otimes \mu'} =
 \sum_\tau \calP_{\lambda \otimes \mu}(\tau) \calP_{\lambda' \otimes \mu'}(\tau) 
 \enspace . \]
 Then~\eqref{eq:prob-hom} implies that the total probability of observing a
homogeneous irrep is 
\begin{multline}
 \label{eq:prob-hom-bound}
\sum_\tau \calP_{\sigma_{\{\lambda,\lambda'\}} \otimes \sigma_{\{\mu,\mu'\}}}(\sigma_{\{\tau,\tau\}}^\pm)
=  \frac{1}{2} \left( \Pcoll_{\lambda \otimes \mu, \lambda' \otimes \mu'}
+  \Pcoll_{\lambda \otimes \mu', \lambda' \otimes \mu} \right) \\
 \le \max \left( \Pcoll_{\lambda \otimes \mu, \lambda' \otimes \mu'},
  \Pcoll_{\lambda \otimes \mu', \lambda' \otimes \mu} \right) \enspace . 
\end{multline}
In the next section, we show that if $\lambda, \mu, \lambda'$ and $\mu'$ are 
typical irreps of $S_n$, then no irrep $\tau$ occurs too often in any of these natural distributions, 
and so the probability of a collision is small.

\section{Collisions, smoothness, and characters}
\label{sec:collisions}

Let us bound the probability $\Pcoll = \Pcoll_{\lambda \otimes \mu, \lambda'
\otimes \mu'}$ that the natural distributions in $\lambda \otimes \mu$ and
$\lambda' \otimes \mu'$ collide.  The idea is that $\Pcoll$ is small as long as
both of either or both of these distributions is \emph{smooth}, in the sense
that they are spread fairly uniformly across many $\tau$.  The following lemmas  
show that this notion of smoothness can be related to bounds on the normalized
characters of these representations.  
First, we present a lemma which relates the natural distribution in a representation $\rho$ 
to the Plancherel distribution.  

\begin{lemma}
\label{lem:nat-planch}
Let $\rho$ be a (possibly reducible) representation of a group $G$, and let 
$\Prho(\tau)$ denote the probability of observing an irrep $\tau \in \wg$ 
under the natural distribution in $\rho$.  Let $X \subseteq \wg$, 
and let $\Prho(X) = \sum_{\tau \in X} \Prho(\tau)$ and 
$\planch(X) = \sum_{\tau \in X} d_\tau^2/|G|$ denote the total probability of 
observing an irrep in $X$ in the natural and Plancherel distributions respectively.  Then 
\[ \Prho(X) \le \sqrt{\planch(X)} \sqrt{\sum_{g \in G} \abs{\frac{\chi_\rho(g)}{d_\rho}}^2} 
\enspace . \]
\end{lemma}
\begin{proof}
In general, we have
\[ \Prho(\tau) = \frac{d_\tau}{d_\rho} \inner{\chi_\tau}{\chi_\rho}_G \enspace . \]
Therefore, if we define 
\[ \one_X = \sum_{\tau \in X} d_\tau \chi_\tau \enspace , \]
then by Cauchy-Schwartz we have
\begin{multline*}
\Prho(X) = \inner{\one_X}{\frac{\chi_\rho}{d_\rho}}_G 
\le \sqrt{\inner{\one_X}{\one_X}_G} \sqrt{\inner{\frac{\chi_\rho}{d_\rho}}{\frac{\chi_\rho}{d_\rho}}_G} 
= \\ \sqrt{\frac{1}{|G|} \inner{\one_X}{\one_X}_G} \sqrt{\sum_{g \in G} \abs{\frac{\chi_\rho(g)}{d_\rho}}^2} 
\end{multline*}
and by Schur's lemma we have
\[ \frac{1}{|G|} \inner{\one_X}{\one_X}_G 
= \sum_{\tau \in X} \frac{d_\tau^2}{|G|} \inner{\chi_\tau}{\chi_\tau}_G 
= \sum_{\tau \in X} \frac{d_\tau^2}{|G|} 
= \planch(X) \]
which completes the proof.
\end{proof}

Now we bound the probability of a collision as follows.
\begin{lemma}
\label{lem:coll}
Given a family of groups $\{G_n\}$, say that an irrep $\lambda$ of $G_n$ is \emph{$f(n)$-smooth} if 
\[ \sum_{g \in G_n} \abs{\frac{\chi_\lambda(g)}{d_\lambda}}^4 \le f(n) \enspace . \]
Suppose that $\lambda$ and $\mu$ are $f(n)$-smooth.  Then
\[ \Pcoll \le \frac{\max_\tau d_\tau}{\sqrt{|G_n|}} \sqrt{f(n)} \enspace . \]
\end{lemma}

\begin{proof}
We write $G$ for $G_n$ to conserve ink.  We have $\Pcoll \le \max_\tau \Plmt$.  Setting $\rho = \lambda \otimes \mu$ and $X=\{\tau\}$ in Lemma~\ref{lem:nat-planch} and applying Cauchy-Schwartz gives
\begin{multline*}
\Pcoll 
\le \sqrt{\max_\tau \planch(\tau)} 
\sqrt{ \sum_{g \in G} \abs{\frac{\chi_\lambda(g)}{d_\lambda}}^2 \abs{\frac{\chi_\mu(g)}{d_\mu}}^2} 
\le \\ \frac{\max_\tau d_\tau}{\sqrt{|G|}}
\sqrt[4]{ \sum_{g \in G} \abs{\frac{\chi_\lambda(g)}{d_\lambda}}^4
\sum_{g \in G} \abs{\frac{\chi_\mu(g)}{d_\mu}}^4} 
\end{multline*}
which completes the proof.
\end{proof}

Now let us focus on the case relevant to Graph Isomorphism, where $G=S_n$.  
Here we recall that each irrep of the symmetric group
$S_n$ corresponds to a \emph{Young diagram}, or equivalently an integer
partition $\lambda_1 \ge \lambda_2 \ge \cdots$ where $\sum_i \lambda_i = n$.  
The maximum dimension of any irrep is bounded by the following result of Vershik
and Kerov: \begin{theorem}[\cite{VershikK85}]
\label{thm:max-dimension}
  There is a constant $\hat{c} > 0$ such that 
$\max_\tau d_\tau \leq e^{-(\hat{c}/2) \sqrt{n}} \sqrt{n!}$.
\end{theorem}
\noindent
In this case, Lemma~\ref{lem:coll} gives 
\begin{equation}\label{eq:coll}
\Pcoll \le e^{-(\hat{c}/2) \sqrt{n}} \sqrt{f(n)} \enspace .
\end{equation}
Therefore, our goal is to show that typical irreps of $S_n$ are $f(n)$-smooth
where $f(n)$ grows slowly enough with $n$, and to show inductively that with high probability 
all the irreps we observe throughout the sieve are typical.  
We do this by defining a typical irrep as follows.

\begin{definition}
\label{def:typical}
Let $D > e$ be a fixed constant, and say that an irrep $\lambda$ of $S_n$ is \emph{typical} 
if the following two conditions hold true:
\begin{itemize}
 \item the height and width of its Young diagram are less than $D\sqrt{n}$ or, in other words,
if the Young diagram is $D$--balanced  \cite{Biane98},

\item the dimension $d_{\lambda}$ fulfills
\[ d_\lambda > e^{-\frac{1}{2} \sqrt{n} \log n} \sqrt{n!} \enspace . \]

\end{itemize}
\end{definition}

\noindent
To motivate this definition, and to provide the base case for our induction, we show the following.

\begin{lemma}
\label{lem:planch-typical}
There are constants $c > 0$ and $n_0$ such that, 
if $\lambda$ has $n$ boxes with $n>n_0$ and $\lambda$
is chosen according to the Plancherel distribution, 
then $\lambda$ is typical with probability at least $1-e^{-c \sqrt{n}}$.
\end{lemma}

\begin{proof}
Firstly, we bound the probability that $\lambda$ is  not $D$-balanced. 
The Robinson-Schensted correspondence~\cite{Fulton97} maps permutations to Young diagrams in such a way that the uniform measure on $S_n$ maps to the Plan\-che\-rel measure.  
In addition, the width (resp.\ height) of the Young diagram is equal to the length of the longest increasing (resp.\ decreasing) subsequence.  Therefore, the probability in the Plancherel measure that an irrep is not typical is at most twice the probability that a random permutation has an increasing subsequence of length $w = D \sqrt{n}$.  

The problem of determining the typical size of the longest increasing subsequence is known as 
Ulam's problem; it can be solved using representation theory~\cite{Kerov03} or by a beautiful 
hydrodynamic argument~\cite{AldousD}, and indeed this Lemma holds even if we take $D > 2$ 
in Definition \ref{def:typical}.  Here we content ourselves with an elementary bound for $D>e$.   
By Markov's inequality, the probability an increasing subsequence of length $w = D \sqrt{n}$ 
is at most the expected number of such subsequences, which is 
\begin{equation}
\label{eq:probabilitysmall-a} 
\binom{n}{w} \frac{1}{w!} < \left( \frac{e^2 n}{w^2} \right)^{\!w} = 
\left( \frac{e^2}{D^2} \right)^{D\sqrt{n}} 
\end{equation}
where we used Stirling's approximation $w! > w^w e^{-w}$.  

Secondly, we shall bound the probability that 
\begin{equation}
\label{eq:small-dimension}
d_\lambda \leq e^{-\frac{1}{2} \sqrt{n} \log n} \sqrt{n!}\enspace.  
\end{equation}
The number of irreps is the partition number
\[p(n) = (1+o(1)) \frac{1}{4\sqrt{3}\cdot n} \,e^{\delta \sqrt{n}} < e^{\delta \sqrt{n}} \]
where 
\[ \delta = \sqrt{2/3} \,\pi \enspace ; \]
therefore the Plancherel measure of the set of irreps $\lambda$ of $S_n$ for which
\eqref{eq:small-dimension} holds true  
is at most the number of irreps times the measure of a single such $\lambda$, 
so this probability is at most
%
%
\begin{equation} 
\label{eq:probabilitysmall-b}
p(n) \frac{d_\lambda^2}{n!} 
< e^{\delta \sqrt{n}} e^{-\sqrt{n} \log n} 
= e^{-\omega(\sqrt{n})} \enspace . 
\end{equation}

The sum of the probabilities \eqref{eq:probabilitysmall-a} and 
\eqref{eq:probabilitysmall-b} is bounded from above by $e^{-c \sqrt{n}}$ for sufficiently 
small $c>0$ and for $n$ sufficiently large.
\end{proof}

Given a permutation $\pi$, let $t(\pi)$ denote the length of the shortest 
sequence of transpositions whose product is $\pi$; for instance, if $\pi$ is a single $k$-cycle, then 
$t(\pi) = k-1$.

\begin{lemma}
\label{lem:large-support}
There is a constant $A$ such that, for $n$ sufficiently large, the normalized character of all typical $\lambda$ obeys
\[ \abs{\frac{\chi_\lambda(\pi)}{d_\lambda}} \le \left(\frac{A}{\sqrt{n}}  \right)^{t(\pi)}  \]
for all $\pi\in S_n$ with $t(\pi) > \sqrt{n} \log n$.
\end{lemma}

\begin{proof}
We use the Murnaghan-Nakayama formula for the character~\cite{JamesK1981}.  A \emph{ribbon tile} of length $k$ is a polyomino of $k$ cells, arranged in a path where each step is up or to the right.  Given a Young diagram $\lambda$ and a permutation $\pi$ with cycle structure $k_1 \ge k_2 \ge \cdots$, a \emph{consistent tiling} consists of removing a ribbon tile of length $k_1$ from the boundary of $\lambda$, then one of length $k_2$, and so on, with the requirement that the remaining part of $\lambda$ is a Young diagram at each step.  Let $h_i$ denote the height of the ribbon tile corresponding to the $i$th cycle: then the Murnaghan-Nakayama formula states that
\begin{equation}
\label{eq:m-n}
 \chi_\lambda(\pi) = \sum_T \prod_i (-1)^{h_i+1} 
\end{equation}
where the sum is over all consistent tilings $T$.  

\begin{figure}[tb]
  \begin{center}
    \includegraphics[width=0.6\columnwidth]{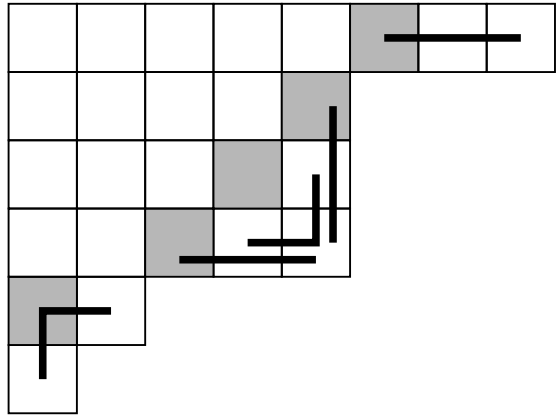}
  \end{center}
\caption{We associate each possible location for a ribbon tile of fixed length $k$ with a cell (shaded) which is above the tile's lower end and to the left of its upper end.  The resulting sequence of cells moves up and to the right at each step, implying that the number of locations is less than $\sqrt{2n}$.  Here $k=3$.}
\label{fig:locations}
\end{figure}

Clearly the number of consistent tilings is an upper bound on $\abs{\chi_\lambda(\pi)}$.  Now, we claim that for any fixed $k$, the number of possible locations for a ribbon tile of length $k$ on the boundary of a Young diagram $\lambda$ of size $n$ is less than $\sqrt{2 n}$.  To see this, associate each one with the cell of $\lambda$ which is directly above the tile's lower end, and directly to the left of its upper end, as shown in Figure~\ref{fig:locations}.  A little thought reveals that the resulting sequence of cells has the property that each one is above and to the right of the previous one.  Therefore, if there are $\ell$ locations, we have
\[ n \ge \sum_{i=1}^\ell i > \ell^2 / 2 \enspace . \]
and so $\ell < \sqrt{2n}$.  It follows that the number of ways to remove the ribbon tiles corresponding to the $c(\pi)$ nontrivial cycles is less than  
\[ (2n)^{c(\pi)/2} \enspace . \]
Moreover, after these ribbon tiles are removed, the number of consistent tilings of the remaining Young diagram is simply the dimension of the corresponding irrep of $S_{n-s(\pi)}$, which is less than  $\sqrt{|S_{n-s(\pi)}|} = \sqrt{(n-s(\pi))!}$.  Therefore, if $\lambda$ is typical we have
\begin{align*} 
\abs{\frac{\chi_\lambda(\pi)}{d_\lambda}} 
&< \frac{(2n)^{c(\pi)/2} \sqrt{(n-s(\pi))!}}{e^{-\sqrt{n} \log n} \sqrt{n!}} \\
&< 2 \cdot e^{\sqrt{n} \log n} \,2^{c(\pi)/2} \,e^{s(\pi)/2} \,n^{(c(\pi)-s(\pi))/2} \\
&\le 2 \cdot e^{\sqrt{n} \log n} \,(\sqrt{2} e)^{t(\pi)} \,n^{-t(\pi)/2} \enspace .
\end{align*}
Here we used the bound $(n-s)!/n! < 4 \cdot n^{-s} e^s$, implied by Stirling's approximation,  
and the facts that $c(\pi) \le t(\pi)$, $s(\pi) \le 2 t(\pi)$, and $t(\pi) = s(\pi) - c(\pi)$.  Finally, if $t(\pi) > \sqrt{n} \log n$, the term $e^{\sqrt{n} \log n}$ can be absorbed into $A^{t(\pi)}$, and Lemma holds for any $A > \sqrt{2} e^2$.
\end{proof}


%

\begin{lemma}[Rattan and \'Sniady \cite{RattanSniady06}]
\label{lem:rattan-sniady}
For every $D>0$ there exists a constant $A'$ with the following property. If
$\lambda$ is a Young diagram with $n$ boxes which has at most $D\sqrt{n}$
rows and columns and $\pi\in S_n$ is a permutation then
\begin{equation} 
 \label{eq:szacowanie}
 \abs{\frac{\chi_\lambda(\pi)}{d_\lambda}}  <
\left( \frac{A' \max(1,t(\pi)^2/n )}{\sqrt{n}} \right)^{t(\pi)}. 
\end{equation}
\end{lemma}

\begin{lemma}
\label{lem:bigsmooth}
All typical irreps $\lambda$ are $O(1)$-smooth.
\end{lemma}
\begin{proof}
If $\lambda$ is typical, then Lemma \ref{lem:large-support} implies
\[ \sum_{\substack{\pi \in S_n\\ t(\pi)>\sqrt{n} \log n}} \abs{\frac{\chi_\lambda(\pi)}{d_\lambda}}^4 
\le \sum_{\pi \in S_n} (A^{t(\pi)} n^{-t(\pi)/2})^4
= \sum_{\pi \in S_n} z^{t(\pi)} \]
for $z=A^4/n^2$. Since each $\pi \in S_n$ appears exactly once in the product
$$ \big[1+ (12) \big] \big[1+ (13)+(23) \big] \cdots \big[1+(1n)+\cdots+(n-1,n)\big] $$
where $(i,j)$ denotes the transposition interchanging $i$ and $j$, 
and since each product of the summands provides a factorization of $\pi$ 
into a minimal number of transpositions, we have 
\begin{multline*} \sum_{\pi \in S_n} z^{t(\pi)}  = (1+z) (1+2z) \cdots (1+(n-
1) z)< \\ e^{z} e^{2z} \cdots e^{(n-1)z} <  e^{z n^2/2} = e^{A^4/2} \end{multline*}
therefore
\begin{equation}
\label{eq:long-permutations-ok}
\sum_{\substack{\pi \in S_n\\ t(\pi)>\sqrt{n} \log n}} \abs{\frac{\chi_\lambda(\pi)}{d_\lambda}}^4 <e^{A^4/2}\enspace.
\end{equation}

Very similar but slightly more involved reasoning can be applied to the estimate from
Lemma \ref{lem:rattan-sniady} (for details we refer to \cite{RattanSniady06}) which shows
that there exist constants $E>0$ and $E'$ (which depend only on $D$)
with a property that if 
a Young diagram $\lambda$ with $n$ boxes has at most $D\sqrt{n}$ boxes in each 
row and column then
\begin{equation}
\label{eq:moore-russell} 
\sum_{\substack{\pi\in S_n,\\ t(\pi)\leq E n^{4/7}}}  \abs{\frac{\chi_\lambda(\pi)}{d_\lambda}}^4 \leq E'\enspace.
\end{equation}

The domains of the summations in the inequalities \eqref{eq:long-permutations-ok} 
and \eqref{eq:moore-russell} cover the whole group $S_n$ for sufficiently large 
$n$  which finishes the proof.
\end{proof}

\begin{lemma}
\label{lem:inductivebig}
There are constants $c' > 0$ and $n_0$ 
such that for all pairs of typical irreps $\lambda$ and $\mu$, if $\tau$ is 
chosen according to the natural distribution $\Plmt$, 
then $\tau$ is typical with probability at least $1-e^{-c' \sqrt{n}}$ if
$n>n_0$.
\end{lemma}
\begin{proof}  Let $X$ be the set of atypical representations, and let $\rho = \lambda \otimes \mu$.  
Then applying Lemma~\ref{lem:nat-planch} and Lemma~\ref{lem:planch-typical}, 
using Cauchy-Schwartz as in the proof of Lemma~\ref{lem:coll}, and finally applying 
Lemma~\ref{lem:bigsmooth} gives 
\begin{align*}
\calP_{\lambda \otimes \rho}(X) 
&\le \sqrt{\planch(X)} 
\sqrt{ \sum_{g \in G} \abs{\frac{\chi_\lambda(g)}{d_\lambda}}^2 \abs{\frac{\chi_\mu(g)}{d_\mu}}^2} \\
&\le e^{-(c/2) \sqrt{n}} 
\sqrt[4]{ \sum_{g \in G} \abs{\frac{\chi_\lambda(g)}{d_\lambda}}^4
\sum_{g \in G} \abs{\frac{\chi_\mu(g)}{d_\mu}}^4} 
\\ & \le e^{-(c/2) \sqrt{n}} \,O(1)
\end{align*}
which completes the proof for any $c' < c/2$.
\end{proof}

\section{Proof of the main result}
\label{sec:main}

We are now in a position to present our main result.

\begin{theorem}
\label{thm:main}
Let $\hat{c}, c, c'$ be the constants defined above. Then 
for any constants $a, b$ such that $a+b < \min(\hat{c}/2,c,c')$, 
no sieve algorithm which combines less than $e^{a \sqrt{n}}$ coset states can solve 
Graph Isomorphism with success probability greater than $e^{-b \sqrt{n}}$. 
\end{theorem}

\begin{proof}
We first consider the behavior of a sieve algorithm $A$ in the case where
the hidden subgroup $H \subset S_n \wr \Z_2$ is trivial.  For convenience,
let us say that a representation $\sigma_{\{\lambda, \mu\}}$ of $S_n \wr \Z_2$ 
is typical if both $\lambda$ and $\mu$ are.  We will establish that with overwhelming 
probability, all the irrep labels observed by $A$ are both typical and inhomogeneous.  

Let $\ell$ be the number of coset states initially generated by 
the algorithm.  We begin by showing that with high probability, the irrep labels
on the $\ell$ leaves, i.e., those resulting from weak Fourier sampling these coset states, are all
both typical and homogeneous.  
If $H$ is trivial, then these irrep labels are
Plancherel-distributed; by~\eqref{eq:planch-wreath-inhom} the probability
that a given one fails to be typical is at most twice the probability that a
Plancherel-distributed irrep of $S_n$ fails to be, which by Lemma~\ref{lem:planch-typical} 
is at most $e^{-c \sqrt{n}}$.  
Moreover, by~\eqref{eq:planch-wreath-hom} the probability that the label of a
given leaf is homogeneous is the probability that we observe the same irrep of
$S_n$ twice in two independent samples of the Plancherel distribution, which using 
Theorem~\ref{thm:max-dimension} is 
\[ \sum_{\lambda} \left(
\frac{d_\lambda^2}{n!} \right)^{\!2}
< \max_\lambda \frac{d_\lambda^2}{n!}
\le e^{-\hat{c} \sqrt{n}} \enspace . \]
Thus the combined probability that any of the $\ell$ leaves have a label which
is not both typical and inhomogeneous is at most
\begin{equation}
\label{eq:event1}
\ell \left( 2e^{-c \sqrt{n}} + e^{-\hat{c} \sqrt{n}} \right) 
\enspace .
\end{equation}

Now, assume inductively that all the irreps observed by the algorithm before the
$i$th combine-and-measure step are typical and inhomogeneous, and that the
$i$th step combines states with two such labels $\sigma_{\{\lambda,\lambda'\}}$
and $\sigma_{\{\mu,\mu'\}}$.  By~\eqref{eq:prob-hom-bound}, the
probability this results in a homogeneous irrep is bounded by the probability
$\Pcoll$ of a collision between a pair of natural distributions in $S_n$.  Then
Theorem~\ref{thm:max-dimension} and Lemmas~\ref{lem:coll} and~\ref{lem:bigsmooth} and 
imply that this probability is bounded by 
\[ \Pcoll \le e^{-(\hat{c}/2) \sqrt{n}} \,O(1) \enspace . \]
In addition, Lemma~\ref{lem:inductivebig} implies that
the the probability the observed irrep fails to be typical is at most
$e^{-c'\sqrt{n}}$.  Since each combine-and-measure step reduces the number
of states by one, there are less than $\ell$ such steps; taking a union bound
over all of them, the probability that any of the observed irreps fail to be
both homogeneous and typical is 
\begin{equation}
\label{eq:event2}
 \ell \left( e^{-(\hat{c}/2) \sqrt{n}} \,O(1) + e^{-c' \sqrt{n}} \right) 
 \enspace .
\end{equation}
Let us call a transcript inhomogeneous if all of its irrep labels are.  
Combining~\eqref{eq:event1} and~\eqref{eq:event2} and setting $\ell < e^{a \sqrt{n}}$, 
we see that, for $n$ sufficiently large, $A$'s transcript is inhomogeneous with probability 
greater than $1-e^{-b \sqrt{n}}$ for any $b < \min(\hat{c}/2, c, c')-a$.
 
Now consider $A$'s behavior in the case of a nontrivial hidden subgroup $H =
\{1,m\}$.  Inductively applying Equation~\eqref{eq:equality} shows that the
probability of observing any inhomogeneous transcript is exactly the same as it
would have been if $H$ were trivial. Thus the total variation distance between
the distribution of transcripts generated by $A$ in these two cases is less than  
$e^{-b \sqrt{n}}$, and the theorem is proved.
\end{proof}

\section*{Acknowledgments}

We are very grateful to Philippe Biane, Persi Diaconis and Valentin F\'eray
for helpful conversations about the character theory of the symmetric groups,
and to Sally Milius, Tracy Conrad and Rosemary Moore for inspiration.  C.M. and A.R. 
are supported by NSF grants CCR-0220070, EIA-0218563, and CCF-0524613, and ARO
contract W911NF-04-R-0009.  
P.\'S. is supported by the MNiSW research grant 1-P03A-013-30 and by
the EC Marie Curie Host Fellowship for the Transfer of Knowledge
``Harmonic Analysis, Nonlinear Analysis and Probability,'' contract
MTKD-CT-2004-013389.

\newcommand{\etalchar}[1]{$^{#1}$}
\ifx \k \undefined \let \k = \c \immediate\write16{Ogonek accent unavailable:
  replaced by cedilla} \fi\ifx \ocirc \undefined \def \ocirc
  #1{{\accent'27#1}}\fi\ifx \mathbb \undefined \def \mathbb #1{{\bf #1}}
  \fi\ifx \mathbb \undefined \def \mathbb #1{{\bf #1}}\fi


\begin{thebibliography}{HMR+06}

\bibitem[AJL06]{Aharonov:2006:PQA}
Dorit Aharonov, Vaughan Jones, and Zeph Landau.
\newblock A polynomial quantum algorithm for approximating the {Jones} polynomial.
\newblock {\em Proc.\ 38th Symposium on Theory of Computing}, 427--436, 2006.

\bibitem[AMR06]{AlagicMR06}
Gorjan {Alagi\'{c}}, Cristopher Moore, and Alexander Russell.
\newblock Quantum algorithms for {S}imon's problem over general groups.
\newblock {\em Proc.\ 18th Symposium on Discrete Algorithms} (2007).

\bibitem[AD95]{AldousD}
David Aldous and Persi Diaconis.
\newblock Hammersley's interacting particle process and longest increasing subsequences. 
\newblock \emph{Probab. Theory Relat. Fields} {\bf 103} (1995), 199--213.

\bibitem[Bab80]{Babai:1980:CCL}
L{\'a}szl{\'o} Babai.
\newblock On the complexity of canonical labeling of strongly regular graphs.
\newblock {\em SIAM J. Computing}, 9(1):212--216, 1980.

\bibitem[Bab83]{Babai:1983:CLG}
L{\'a}szl{\'o} Babai and Eugene~M. Luks.
\newblock Canonical labeling of graphs.
\newblock {\em Proc.\ 15th Symposium on Theory of Computing}, 171--183.

\bibitem[BCvD05]{Bacon:2005:OMEA}
David Bacon, Andrew Childs, and Wim {van Dam}.
\newblock From optimal measurement to efficient quantum algorithms for the
  hidden subgroup problem over semidirect product groups.
\newblock {\em Proc.\ 46th Symposium on Foundations of Computer Science}, 469--478, 2005.

\bibitem[BCvD06]{BaconCvD}
David Bacon, Andrew Childs, and Wim van Dam.
\newblock Optimal measurements for the dihedral hidden subgroup problem.
\newblock {\em Chicago Journal of Theoretical Computer Science}, 2006.

\bibitem[Bia98]{Biane98}
Philippe Biane.
\newblock Representations of symmetric groups and free probability.
\newblock {\em Advances in Mathematics}, 138(1):126--181, 1998.

\bibitem[FIM{\etalchar{+}}03]{Friedl:2003:HTO}
Katalin Friedl, G{\'a}bor Ivanyos, Fr{\'e}d{\'e}ric Magniez, Miklos Santha, and Pranab Sen.
\newblock Hidden translation and orbit coset in quantum computing.
\newblock {\em Proc.\ 35th Symposium on Theory of Computing}, 1--9, 2003.

\bibitem[Ful97]{Fulton97}
William Fulton.
\newblock {\em Young Tableaux: with Applications to Representation Theory and
  Geometry}, volume~35 of {\em Student Texts}.
\newblock London Mathematical Society, 1997.

\bibitem[GSVV01]{Grigni:2001:QMA}
Michelangelo Grigni, Leonard Schulman, Monica Vazirani, and Umesh Vazirani.
\newblock Quantum mechanical algorithms for the nonabelian hidden subgroup problem.
\newblock {\em Proc.\ 33rd Symposium on Theory of Computing}, 68--74, 2001.

\bibitem[Hal02]{Hallgren:2002:PTQ}
Sean Hallgren.
\newblock Polynomial-time quantum algorithms for {Pell}'s equation and the principal ideal problem.
\newblock {\em Proc.\ 34th Symposium on Theory of Computing}, 653--658.

\bibitem[Hal05]{Hallgren:2005:FQA}
Sean Hallgren.
\newblock Fast quantum algorithms for computing the unit group and class group of a number field.
\newblock {\em Proc.\ 37th Symposium on Theory of Computing}, 468--474, 2005.

\bibitem[HMR{\etalchar{+}}06]{Hallgren:2006:LQC}
Sean Hallgren, Cristopher Moore, Martin R{\"o}tteler, Alexander Russell, and Pranab Sen.
\newblock Limitations of quantum coset states for graph isomorphism.
\newblock {\em Proc.\ 38th Symposium on Theory of Computing}, 604--617, 2006.

\bibitem[HRTS00]{Hallgren:2000:NSR}
Sean Hallgren, Alexander Russell, and Amnon Ta-Shma.
\newblock Normal subgroup reconstruction and quantum computation using group representations.
\newblock {\em Proc.\ 32nd  Symposium on Theory of Computing}, 627--635, 2000.

\bibitem[JK81]{JamesK1981}
Gordon James and Adalbert Kerber.
\newblock {\em The representation theory of the symmetric group}, volume~16 of
  {\em Encyclopedia of mathematics and its applications}.
\newblock Addison--Wesley, 1981.

\bibitem[Ker03]{Kerov03}
{S. V.} Kerov.
\newblock {\em Asymptotic representation theory of the symmetric group and its
  applications in analysis}, volume 219 of {\em Translations of Mathematical
  Monographs}.
\newblock American Mathematical Society, 2003.
\newblock Translated by {N. V.} Tsilevich.

\bibitem[Kup05]{Kuperberg}
Greg Kuperberg.
\newblock A subexponential-time quantum algorithm for the dihedral hidden subgroup.
\newblock {\em SIAM J. Computing} 35(1): 170--188, 2005.

\bibitem[MR05a]{MooreR:banff}
Cristopher Moore and Alexander Russell.
\newblock Explicit multiregister measurements for hidden subgroup problems; or,
  {Fourier} sampling strikes back.
\newblock Preprint, quant-ph/0504067 (2005).

\bibitem[MR05b]{MooreR:part2}
Cristopher Moore and Alexander Russell.
\newblock The symmetric group defies strong {Fourier} sampling: part {II}.
\newblock Preprint, quant-ph/0501066 (2005).

\bibitem[MR06]{MooreR:sieve}
Cristopher Moore and Alexander Russell.
\newblock On the impossibility of a quantum sieve algorithm for {Graph} {Isomorphism}.
\newblock Preprint, quant-ph/0609138 (2006).

\bibitem[MRS04]{SODA::MooreRS2004}
Cristopher Moore, Alexander Russell, and Leonard Schulman.
\newblock The power of basis selection in fourier sampling: hidden subgroup problems in affine groups.
\newblock {\em Proc.\ 15th Symposium on Discrete Algorithms}, 1113--1122, 2004.

\bibitem[MRS05]{Moore:2005:SGF}
Cristopher Moore, Alexander Russell, and Leonard Schulman.
\newblock The symmetric group defies {Fourier} sampling.
\newblock {\em Proc.\ 46th Symposium on Foundations of Computer Science}, 479--488, 2005.

\bibitem[R\'S06]{RattanSniady06}
Amarpreet Rattan and Piotr \'Sniady.
\newblock  Upper bound on the characters of the symmetric groups for balanced {Young} diagrams and a generalized {Frobenius} formula.
\newblock Preprint, math.RT/0610540 (2006).

\bibitem[Reg02]{Regev:2002:QCL}
Oded Regev.
\newblock Quantum computation and lattice problems.
\newblock {\em Proc.\ 43rd Symposium on Foundations of Computer Science}, 520--529, 2002.

\bibitem[Ser77]{Serre77}
Jean-Pierre Serre.
\newblock {\em Linear Representations of Finite Groups}.
\newblock Number~42 in Graduate Texts in Mathematics. Springer-Verlag, 1977.

\bibitem[Sho94]{Shor:1994:AQC}
Peter W. Shor.
\newblock Algorithms for quantum computation: discrete logarithms and factoring.
\newblock {\em Proc.\ 35th Symposium on Foundations of Computer Science}, 124--134, 1994.

\bibitem[Sim94]{Simon:1994:PQC}
D. R. Simon.
\newblock On the power of quantum computation.
\newblock {\em Proc.\ 35th Symposium on Foundations of Computer Science}, 116--123, 1994.

\bibitem[Spi96]{Spielman}
Daniel A. Spielman. 
\newblock Faster isomorphism testing of strongly regular graphs. 
\newblock {\em Proc.\ 28th Symposium on Theory of Computing}, 576--584, 1996.

\bibitem[VK85]{VershikK85}
{A. M.} Vershik and {S. V.} Kerov.
\newblock Asymptotic behavior of the maximum and generic dimensions of
  irreducible representations of the symmetric group.
\newblock {\em Funk. Anal. i Prolizhen}, 19(1):25--36, 1985.
\newblock English translation, \emph{Funct. Anal. Appl.} 19(1):21--31, 1989.

\end{thebibliography}
\end{document}